\documentclass[sigconf]{acmart}

\usepackage{graphicx} 
\usepackage{float}
\usepackage{booktabs} 
\usepackage{enumitem,kantlipsum}
\usepackage{threeparttable}
\usepackage{algorithm}
\usepackage{algpseudocode}
\usepackage{multirow}
\usepackage{array}
\usepackage{booktabs}
\usepackage{subfig}

\begin{document}
\copyrightyear{2019}
\acmYear{2019}
\setcopyright{acmcopyright}
\acmConference[SIGIR '19]{Proceedings of the 42nd International ACM SIGIR Conference on Research and Development in Information Retrieval}{July 21--25, 2019}{Paris, France}
\acmBooktitle{Proceedings of the 42nd International ACM SIGIR Conference on Research and Development in Information Retrieval (SIGIR '19), July 21--25, 2019, Paris, France}
\acmPrice{15.00}
\acmDOI{10.1145/3331184.3331188} \acmISBN{978-1-4503-6172-9/19/07}

\settopmatter{printacmref=true}
\fancyhead{}

\title{Relational Collaborative Filtering: \\Modeling Multiple Item Relations for Recommendation}

\author{Xin Xin}
\affiliation{%
  \institution{School of Computing Science}
  \institution{University of Glasgow}
}
\email{x.xin.1@research.gla.ac.uk}

\author{Xiangnan He}
\affiliation{%
  \institution{School of Information Science and Technology, USTC}
}
\email{xiangnanhe@gmail.com}

\author{Yongfeng Zhang}
\affiliation{%
  \institution{Department of Computer Science}
  \institution{Rutgers University}
}
\email{yongfeng.zhang@rutgers.edu}

\author{Yongdong Zhang}
\affiliation{%
  \institution{School of Information Science and Technology, USTC}
}
\email{zhyd73@ustc.edu.cn}

\author{Joemon Jose}
\affiliation{%
  \institution{School of Computing Science}
  \institution{University of Glasgow}
}
\email{Joemon.Jose@glasgow.ac.uk}


\begin{abstract}
Existing item-based collaborative filtering (ICF) methods leverage only the relation of \textit{collaborative similarity} --- i.e., the item similarity evidenced by user interactions like ratings and purchases. 
Nevertheless, there exist multiple relations between items in real-world scenarios, e.g., two movies share the same director, 
two products complement with each other, etc. Distinct from the collaborative similarity that implies co-interact patterns from the user's perspective, these relations reveal fine-grained knowledge on items from different perspectives of meta-data, functionality, etc. However, how to incorporate multiple item relations is less explored in recommendation research. 

In this work, we propose \textit{Relational Collaborative Filtering} (RCF) to exploit multiple item relations in recommender systems. We find that both the relation \textit{type} (e.g., shared director) and the relation \textit{value} (e.g., Steven Spielberg) are crucial in inferring user preference. 
To this end, we develop a two-level hierarchical attention mechanism to model user preference --- the first-level attention discriminates which types of relations are more important, and the second-level attention considers the specific relation values to estimate the contribution of a historical item. 
To make the item embeddings be reflective of the relational structure between items, we further formulate a task to preserve the item relations, and jointly train it with user preference modeling. 
Empirical results on two real datasets demonstrate the strong performance of RCF\footnote{Codes are available at https://github.com/XinGla/RCF.}.
Furthermore, we also conduct qualitative analyses to show the benefits of explanations brought by RCF's modeling of multiple item relations.


\end{abstract}
%
%


\maketitle

\section{Introduction}
Recommender system has been widely deployed in Web applications to address the information overload issue, such as E-commerce platforms, news portals, lifestyle apps, etc. It not only can facilitate the information-seeking process of users, but also can increase the traffic and bring profits to the service provider~\cite{adomavicius2005toward}. 
Among the various recommendation methods, item-based collaborative filtering (ICF) stands out owing to its interpretability and effectiveness~\cite{FISM,he2018nais}, being highly preferred in industrial applications~\cite{Smith:2017,YoutubeRS,pixie}. 
The key assumption of ICF is that a user shall prefer the items that are similar to her historically interacted items~\cite{itemCF,xue2018deepICF,NGCF}. The similarity is typically judged from user interactions --- how likely two items are co-interacted by users in the past. 

Despite prevalence and effectiveness, we argue that existing ICF methods are insufficient, since they only consider the \textit{collaborative similarity} relation, which is macro-level, coarse-grained and lacks of concrete semantics. 
In real-world applications, there typically exist multiple relations between items that have concrete semantics, and they are particularly helpful to understand user behaviors. For example, in the movie domain, some movies may share the same director, actors, or other attributes; in E-commerce, some products may have the same functionality, similar image, etc. These relations reflect the similarity of items from different perspectives, and more importantly, they could affect the decisions of different users differently. For example, after two users ($u1$ and $u2$) watch the same movie ``E.T. the Extra-Terrestrial'', $u1$ likes the director and chooses ``Schindler's List'' to watch next, while $u2$ likes the fiction theme and watches ``The Avenger'' in the next. Without explicitly modeling such micro-level and fine-grained relations between items, it is conceptually difficult to reveal the true reasons behind a user's decision, not to mention to recommend desired items with persuasive explanations like \textit{``The Avenger'' is recommended to you because it is a fiction movie like ``E.T. the Extra-Terrestrial'' you watched before} for the user $u2$.

In this paper, we propose a novel ICF framework \emph{Relational Collaborative Filtering} (RCF), aiming to integrate multiple item relations for better recommendation. To retain the fine-grained semantics of a relation and facilitate the reasoning on user preference, we represent a relation as a concept with a two-level hierarchy:
\begin{enumerate}[leftmargin=*]
\item Relation \emph{type}, which can be shared director and genre in the above movie example, or functionality and visually similar for E-commerce products. It describes how items are related with each other in an abstract way. The collaborative similarity is also a relation type from the macro view of user behaviors. 
\item Relation \emph{value}, which gives details on the shared relation of two items. For example, the value of relation shared director for  ``E.T. the Extra-Terrestrial'' and ``Schindler's List'' is Steven Spielberg, and the values for relation shared genre include fiction, action, romantic, etc. The relation values provide important clues for scrutinizing a user's preference, since a user could weigh different values of a relation type differently when making decisions. 
\end{enumerate}

\begin{figure}
	\centering
    \includegraphics[width=0.48\textwidth]{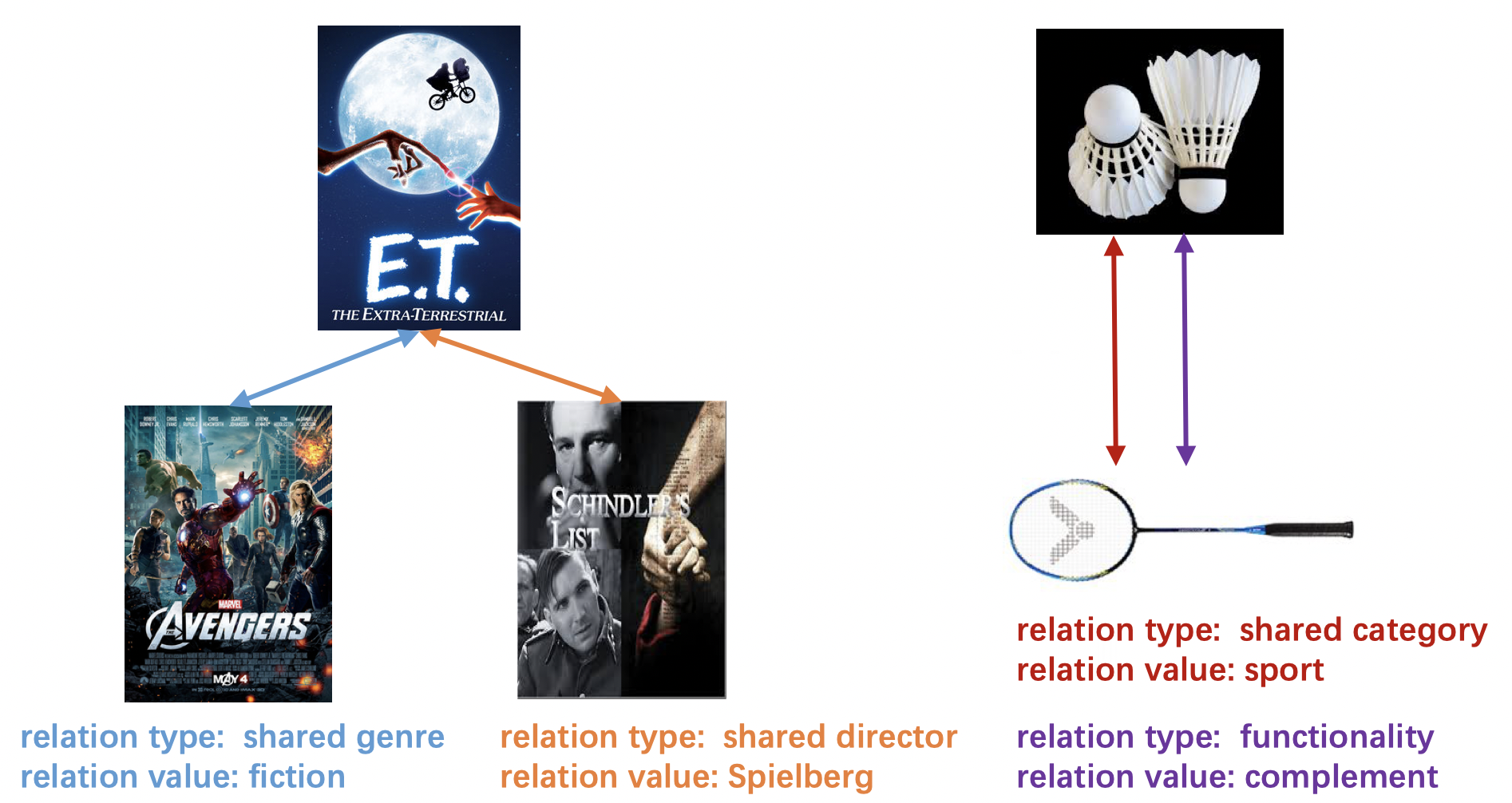}
    \caption{An example of multiple item relations. Each relation is described with a two-level hierarchy of type and value. Multiple relations may exist between two items and the same value may occur in relations of different types. 
    }\vspace{-10pt}
    \label{fig:relation-example}
\end{figure}

Figure \ref{fig:relation-example} gives an illustrative example on the item relations. Note that multiple relations may exist between two items; for example, badminton birdies balls and badminton rackets have two relations of complementary functionality and shared category. Moreover, a relation value may occur in multiple relations of different types; for example, a director can also be the leading actor of other movies, thus it is likely that two types of relations have the same value which refers to the same stuff. When designing a method to handle multiple item relations, these factors should be taken into account, making the problem more complicated than the standard ICF. 

To integrate such relational data into ICF, we devise a two-level neural attention mechanism~\cite{attnention} to model the historically interacted items. Specifically, to predict a user's preference on a target item, the first-level attention examines the types of the relations that connect the interacted items with the target item, and discriminates which types affect more on the user. The second-level attention is operated on the interacted items under each relation type, so as to estimate the contribution of an interacted item in recommending the target item. The two-level attention outputs a weight for each interacted item, which is used to aggregate the embeddings of all interacted items to obtain the user's representation. Furthermore, to enhance the item embeddings with the multi-relational data, we formulate another learning task that preserves the item relations with embedding operations. Finally, we jointly optimize the two tasks to make maximum usage of multiple relations between items. 


To summarize, this work makes the key contributions as follows:
\begin{itemize}
    \item We propose a new and general recommendation task, that is, incorporating the multiple relations between items to better predict user preference. 
    \item We devise a new method RCF, which leverages the relations in two ways:  constructing user embeddings by improved modeling of historically interacted items, and enhancing item embeddings by preserving the relational structure. 
    \item We conduct experiments on two datasets to validate our proposal. Quantitative results show RCF outperforms several recently proposed methods, and qualitative analyses demonstrate the recommendation explanations of RCF with multiple item relations. 
\end{itemize}
\section{Methodology}
We first introduce the problem of using multiple item relations for CF, and then elaborate our proposed RCF method. 

\subsection{Problem Formulation}
\begin{figure}
	\centering
    \setlength{\abovecaptionskip}{-0cm}
    \includegraphics[width=0.48\textwidth]{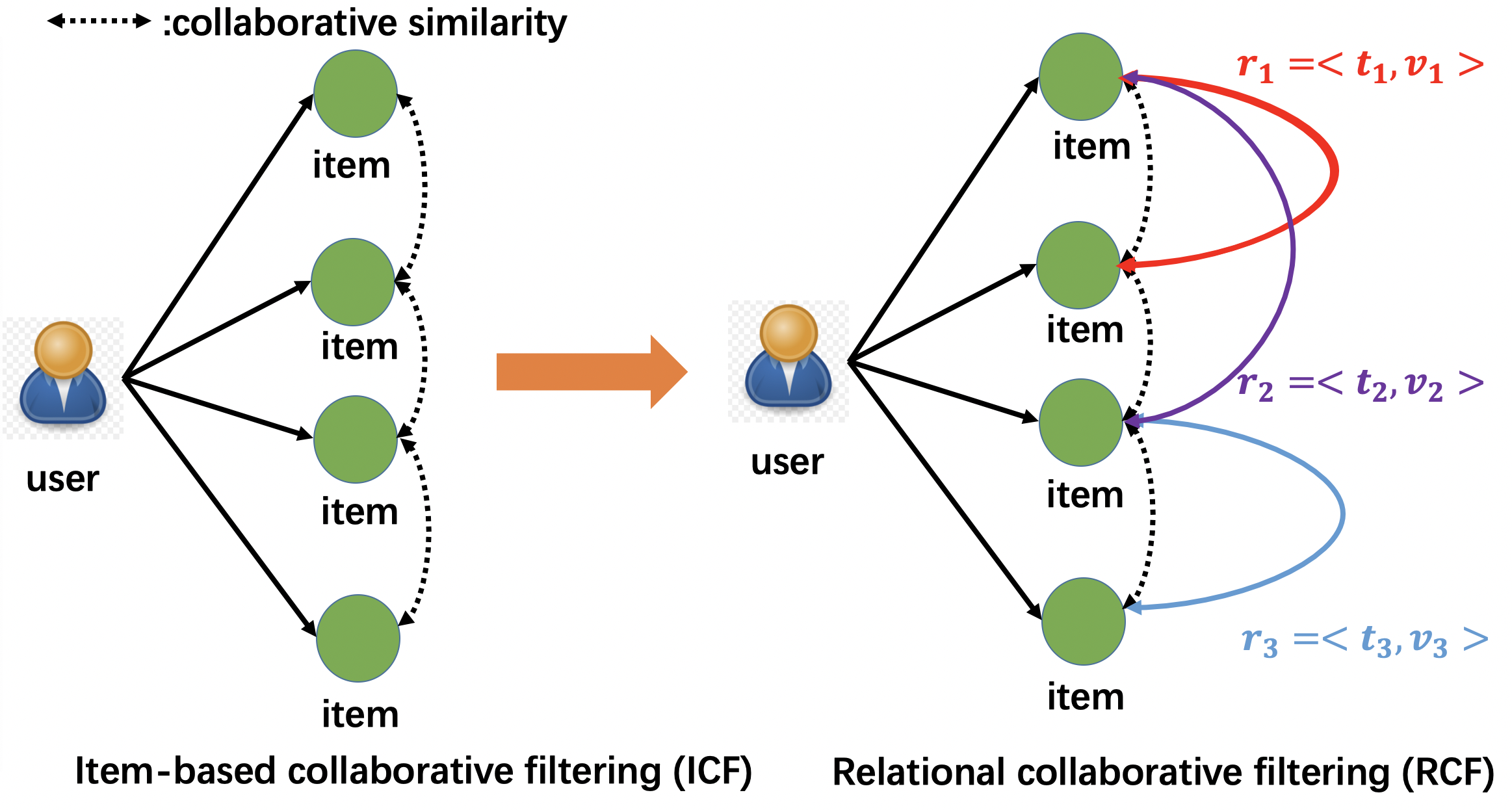}
    \caption{Comparison between ICF and RCF. The links between items of ICF are implicit and single, which denote the collaborative similarity. However, the links between items of RCF are explicit and multiple.} \vspace{-5pt} 
    \label{icf-rcf}
\end{figure}

Given a user and his interaction history, conventional ICF methods aim at generating recommendations based on the collaborative similarity which encode the co-interact patterns of items. Its interaction graph can be shown as the left part of Figure \ref{icf-rcf}, where the links between items are just the implicit collaborative similarity. However, there are multiple item relations in the real world which have meaningful semantics. In this work, we define the item relations as:
\theoremstyle{definition}
\begin{definition}
Given an item pair $(i,j)$, the relations between them are defined as a set of $r=<t,v>$ where $t$ denotes the relation type and $v$ is the relation value. 
\end{definition}
The target of RCF is to generate recommendations based on both the user-item interaction history and item relational data. Generally speaking, the links between items in the interaction graph of RCF contain not only the implicit collaborative similarity, but also the explicit multiple item relations, which are represented by the heterogeneous edges in the right part of Figure \ref{icf-rcf}.
Our notations are summarized in Table \ref{Notations}.
\begin{table}[t]
 \centering
 \setlength{\abovecaptionskip}{-0.05cm}
 \begin{threeparttable}
 \caption{\label{Notations}Notations}
  \begin{tabular}{p{1.3cm}p{6.5cm}}
  \toprule
  Notation & Description\cr
  \midrule
  $\mathcal{U,I}$&the set of users and items\cr
  $\mathcal{T}$&the set of relation types\cr
  $\mathcal{V}$&the set of relation values\cr
  $\mathcal{I}_u^+$& the item set which user $u$ has interacted with\cr
  $\mathcal{I}_{u,i}^t$ & the items in $\mathcal{I}_u^+$ that have the relation of type $t$ with the target item $i$ \cr
  
  %
  $I_r(i,j)$& an indicator function where $I_r(i,j)=1$ if relation $r$ holds for item $i$ and $j$, otherwise 0\cr
  $\mathbf{p}_u\in\mathbb{R}^d$&the ID embedding for user $u \in\mathcal{U}$, which represents the user's inherent interests\cr
  $\mathbf{q}_i\in\mathbb{R}^d$&the embedding for item $i \in\mathcal{I}$ \cr
  $\mathbf{x}_t\in\mathbb{R}^d$&the embedding for relation type $t \in\mathcal{T}$ \cr
  $\mathbf{z}_v\in\mathbb{R}^d$&the embedding for relation value $v \in\mathcal{V}$\cr
  \bottomrule
  \end{tabular}
 \end{threeparttable}
\end{table}

\vspace{+5pt}
\noindent In the remainder of this section, we first present the attention-based model to infer user-item preference. We then illustrate how to model the item relational data to introduce the relational structure between item embeddings. Based on that, we propose to integrate the two parts in an end-to-end fashion through a multi-task learning framework. Finally, we provide a discussion on the relationship between RCF and some other models.
\subsection{User-Item Preference Modeling}
\begin{figure}
	\centering
    \includegraphics[width=0.45\textwidth]{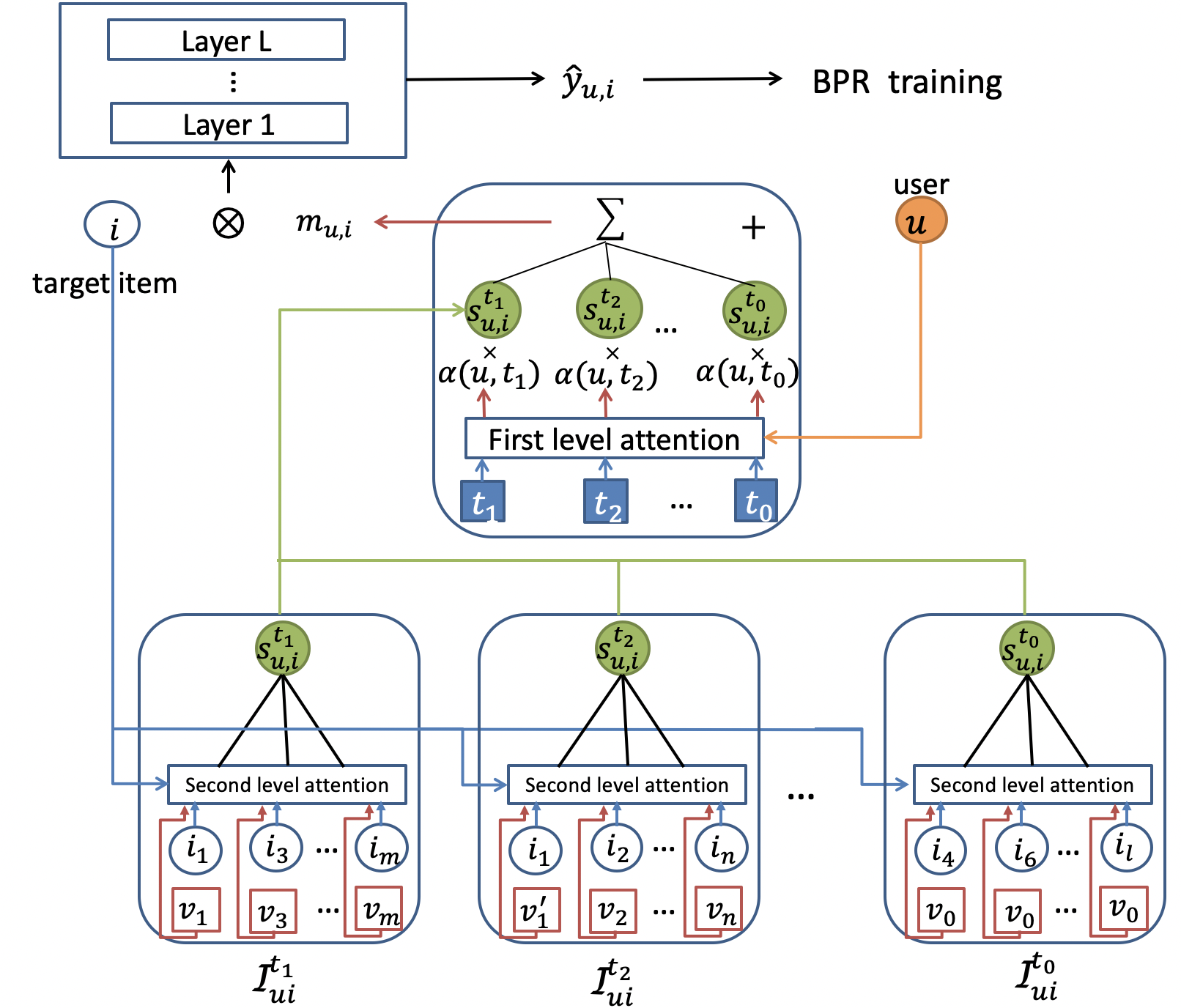}
    \caption{Illustration of the proposed recommendation model. The target-aware user embedding ($\mathbf{m}_{u,i}$) is modeled with a two-level hierarchy attention mechanism. The input of the first level attention contains the user ID embedding and relation types. The second level attention is used to calculate the weights of specific historical items. There are three inputs during this state, including the target item, the historical item and the relation value. Note that one historical item (e.g., $i_1$) can occur in different $\mathcal{I}_{u,i}^t$ when there are multiple relations between it and the target item.} 
    \label{recommendation model}
\end{figure}

An intuitive motivation when modeling user preference is that users tend to pay different weights to relations of different types (e.g., some users may prefer movies which share same actors, some users may prefer movies fall into same genres). Given multiple item relations which consist of relation types and relation values, we propose to use a hierarchy attention mechanism to model the user preference. Figure \ref{recommendation model} demonstrates the overall structure of our model.

Given the item relational data, we first divide the interacted items of user $u$ (i.e., $\mathcal{I}_u^+$) into different sets (i.e., $\mathcal{I}_{u,i}^t$) according to the relation types between these items and the target item. Note that a single item may occur in different $\mathcal{I}_{u,i}^t$ when there are multiple relations between this item and $i$.
Besides, there may be some items which have no explicit relation with the target item. To tackle with these items, we introduce a latent relation $r_0=<t_0,v_0>$ and put these items into $\mathcal{I}_{u,i}^{t_0}$, as shown in Figure \ref{recommendation model}. Here $r_0$ can be regarded as the collaborative similarity which just indicates the item co-interact patterns. Then the target-aware user embedding can be formulated as
\begin{equation}
	\label{mui}
	\mathbf{m}_{u,i} =\mathbf{p}_u+\sum_{t \in \mathcal{T}}\alpha(u,t) \cdot \mathbf{s}_{u,i}^t \hspace{0.1cm},
\end{equation}
where $\alpha(u,t)$ is the first-level attention which aims to calculate the importance of different relation types for this user and $\mathbf{s}_{u,i}^{t}$ describes the user's profile based on the items in $\mathcal{I}_{u,i}^t$. 
More precisely, we define $\alpha(u,t)$ with the standard $softmax$ function:
\begin{equation}
	\label{first level attention}
	\alpha(u,t) =
	\frac{exp(a(\mathbf{p}_u,\mathbf{x}_t))}
	{\sum_{t' \in \mathcal{T}}exp(a(\mathbf{p}_u,\mathbf{x}_{t'}))} \hspace{0.1cm},
\end{equation}
where $a(\mathbf{p}_u,\mathbf{x}_t)$ is the attention score between user $u$ and relation type $t$. We define it with a feedforward neural network, as shown in Eq.(\ref{first level attention score})
\begin{equation}
	\label{first level attention score}
	a(\mathbf{p}_u,\mathbf{x}_t)=\mathbf{h}_1^T(ReLU(\mathbf{W}_1(\mathbf{p}_u \otimes \mathbf{x}_t)+\mathbf{b}_1)).
\end{equation}
$\mathbf{W}_1$ and $\mathbf{b_1}$ are corresponding weight matrix and bias
vector that project the input into a hidden state, and $\mathbf{h}^T_1$ is
the vector which projects the hidden state into the attention score. We term the size of hidden state as ``attention factor'', for which a larger value brings a stronger
representation power for the attention network. $\otimes$ denotes the element-wise product.

The next step is to model $\mathbf{s}_{u,i}^t$. It's obvious that the relation value accounts for an important part during this process. For example, a user may pay attention to genres when watching a movie.  However, among all the genres, he is most interested in fiction other than romantic. As a result, we should consider both the items and the corresponding relation values when modeling the second-level attentions.
From that view, we define $\mathbf{s}_{u,i}^t$ as
\begin{equation}
	\label{suit}
	\mathbf{s}_{u,i}^t =\sum_{j \in \mathcal{I}_{u,i}^t} \beta_t(i,j,v)\cdot \mathbf{q}_{j} \hspace{0.1cm},
\end{equation}
where $\beta_t(i,j,v)$ represents the specific weight of item $j$. 

Similar to Eq.(\ref{first level attention}), a straight-forward solution to calculate $\beta_t(i,j,v)$ is to use the $softmax$ function. However we found that such a simple solution would lead to bad performance. Same observations can also be found in \cite{he2018nais} under similar circumstances. The reason is that the number of items between different $\mathcal{I}_{u,i}^t$ vary greatly. For those items in large $\mathcal{I}_{u,i}^t$, the standard $softmax$ function will have very big denominator, causing the gradient vanishing problem of corresponding $\mathbf{q}_{j}$.

To tackle with this problem, we utilize a $smoothed$ $softmax$ function to replace the standard solution. As a result, the weight $\beta_t(i,j,v)$ is formulated as
\begin{equation}
    \label{second level attention}
	\beta_t(i,j,v)=\frac
	{exp(b_t(\mathbf{q}_i,\mathbf{q}_{j},\mathbf{z}_v))}
	{[\sum_{j' \in \mathcal{I}_{u,i}^t}
	exp(b_t(\mathbf{q}_i,\mathbf{q}_{j'},\mathbf{z}_{v'}]^\rho} \hspace{0.1cm},
\end{equation}
where $\rho$ is a smoothing factor between (0,1] and is commonly set as 0.5 \cite{he2018nais}. $b_t(\mathbf{q}_i,\mathbf{q}_{j},\mathbf{z}_v)$ is the second-level attention score which is defined as
\begin{equation}
	\label{second level attention score}
	b_t(\mathbf{q}_i,\mathbf{q}_j,\mathbf{z}_v)
	= \mathbf{h}_{2,t}^T(ReLU(\mathbf{W}_{2,t}
	\begin{bmatrix} \mathbf{q}_i\\
    				\mathbf{q}_{j}  \\
                    \mathbf{z}_v
    \end{bmatrix}
+\mathbf{b}_{2,t})),
\end{equation}
where $[\cdot]$ denotes the vector concatenation. $\mathbf{W}_{2,t}$, $\mathbf{b}_{2,t}$ and $\mathbf{h}_{2,t}$ are corresponding attention parameters.
Different from Eq.(\ref{first level attention score}) which utilizes element-wise product to learn signals from inputs, here we concatenate the input embeddings and send it to a feedforward neural netwrok. The reason is that there are three inputs when modeling the second-level attentions. Utilizing element-wise product under such situation would have a high risk of suffering from vanishing or exploding gradients.

Now we have completed the modeling of the target-aware user embedding $\mathbf{m}_{u,i}$. Based on that, we utilize a multilayer perceptron (MLP) to calculate the final predicted score of user $u$ on item $i$, which is shown as:\footnote{We introduce a dropout layer \cite{srivastava2014dropout} before each layer of the MLP to prevent overfitting.}
\begin{equation}
	\label{predicted score}
	\hat{y}_{ui}=MLP(\mathbf{m}_{u,i}\otimes \mathbf{q}_i),
\end{equation}
Given the final predicted score $\hat{y}_{ui}$, we want the positive items to have a higher rank than negative ones. We utilize the BPR pairwise learning framework \cite{bpr} to define the objective function, which is shown as
\begin{equation}
	\label{recommendation loss}
	L_{rec} =- \sum_{(u,i,k) \in \mathcal{D}_I}\ln \sigma(\hat{y}_{ui}-\hat{y}_{uk}),
\end{equation}
where $\sigma$ denotes the sigmoid function and $\mathcal{D}_I$ is the set of training triplets:
\begin{equation}
	\label{recommendation data}
	\mathcal{D}_{I} = \left\{(u,i,k)| u \in \mathcal{U}\wedge i \in \mathcal{I}_u^+ \wedge k \in \mathcal{I} \backslash \mathcal{I}_u^+ \right\}.
\end{equation}

\subsection{Item-Item Relational Data Modeling}
The second task of RCF is to model the item relational data. 
Typically, the relational data is organized as knowledge graphs (KG). A knowledge graph is a directed heterogeneous graph in which nodes correspond to entities and edges correspond to relations. It can be represented by a set of triplets $(e_1,r,e_2)$ where $e_1$ denotes the head entity, $r$ is the relation and $e_2$ represents the tail entity. Knowledge graph embedding (KGE) is a popular approach to learn signals from relational data which aims at embedding a knowledge graph into a continuous vector space.

However, directly using techniques from KGE \cite{TransE,TransR,Distmul} to model the item relations of RCF is infeasible due to the following challenges in our specific domain:
\begin{enumerate}[leftmargin=*]
\item The item relation is defined with a two-level hierarchy: relation type and relation value. As shown in Figure \ref{fig:relation-example}, the relation between ``E.T. the Extra-Terrestrial'' and ``The Avenger'' is described as <shared genre,fiction>. To represent this relation properly, we must consider both the first-level (i.e., shared genre) for type constrains and the second-level (i.e., fiction) for model fidelity.  As a result, we can not assign a single embedding for an item relation $r=<t,v>$, which is a common case in the field of KGE \cite{TransE, TransR,Distmul}.
\item Different from the conventional KG which is represented as a directed graph, the item relations are reversible (i.e., the relation $r$ holds for both $(e_1,e_2)$ and $(e_2,e_1)$), resulting in an undirected graph structure. Traditional KGE methods \cite{TransE,TransR} may encounter difficulties under such situations. For example, the most popular TransE \cite{TransE} models the relation between two entities as a translation operation between their embeddings, that is, $\mathbf{e}_1+\mathbf{r}\approx \mathbf{e}_2$ when $(e_1, r, e_2)$ holds, where $\mathbf{e}_1,\mathbf{r},\mathbf{e_2}$ are corresponding embeddings for head entity, relation and tail entity. Based on that, TransE defines the scoring function for this triplet as $f(e_1,r,e_2)=\|\mathbf{e}_1+\mathbf{r}-\mathbf{e}_2\|_2$ where $\|\cdot\|_2$ denotes the $L_2$ norm of a vector. However, because of the undirected structure, we will get both $\mathbf{e}_1+\mathbf{r}\approx \mathbf{e}_2$ and $\mathbf{e}_2+\mathbf{r}\approx \mathbf{e}_1$ on our item relational data.  Optimizing objective functions based on such equations may lead to a trivial solution that $\mathbf{r}\approx \mathbf{0}$ and $\mathbf{e}_1\approx \mathbf{e}_2$.
\end{enumerate}

To tackle with the first challenge, we use the summation of the two-level hierarchy components as relation embeddings.
More precisely, the representation of a specific relation $r=<t,v>$ is formulated as the following equation:
\begin{equation}
\label{relation embedding }
    \mathbf{r}=\mathbf{x}_t+\mathbf{z}_v.
\end{equation}
By doing so, we can make sure that relations with the same type keep similar with each other in some degree. Meanwhile, the model fidelity is also guaranteed because of the value embedding. It also empowers the model with the ability to tackle the situation that same values occur in relations of different types.

To address the second challenge, we find that the source of the trivial solution is the minus operation in TransE, which only suits for directed structures. To model undirected graphs, we need the model which satisfies the commutative law (i.e., $f(e_1,r,e_2)=f(e_2,r,e_1)$). Another state-of-the-art methods of KGE is DistMult \cite{Distmul}. It defines the scoring function as $f(e_1,r,e_2)=\mathbf{e}_1^T\mathbf{M}_r\mathbf{e}_2$, where $\mathbf{M}_r$ is a matrix representation of $r$. It's obvious that DistMult is based on the multiply operation and satisfies the desired commutative property. Based on that, given a triplet $(i,r,j)$ which means item $i$ and $j$ has relation $r$, we define the scoring function for this triplet as
\begin{equation}
\label{relation score}
    f(i,r,j)=\mathbf{q}_i^T \cdot diag(\mathbf{r})\cdot \mathbf{q}_j.
\end{equation}
Here $diag(\mathbf{r})$ denotes a diagonal matrix whose diagonal elements equal to $\mathbf{r}$ correspondingly. 

Similar to the BPR loss used in the recommendation part, we want to maximize  $f(i,r,j)$ for positive examples and minimize it for negative ones. Based on that, the objective function is defined by contrasting the scores of observed triplets $(i,r,j)$ versus unobserved ones $(i,r,j^-)$:
\begin{equation}
	\label{relation loss}
	L_{rel} =- \sum_{(i,r,j,j^{-}) \in \mathcal{D}_R}\ln \sigma(f(i,r,j)-f(i,r, j^{-})),
\end{equation}
where $\mathcal{D}_R$ is defined as
\begin{equation}
	\label{relation data}
	\mathcal{D}_{R} = \left\{(i,r,j,j^-)| i,j,j^{-}\in \mathcal{I} \wedge I_r(i,j)=1 \wedge I_r(i,j^{-})=0 \wedge r \neq r_0\right\}.
\end{equation}
The above objective function encourages the positive item $j$ to be ranked higher than negative items $j^-$ given the context of the head item $i$ and relation $r$. Because $r_0$ is defined as a latent relation so we don't include it during this process. 
\subsection{Multi-Task Learning}
To effectively learn parameters for recommendation, as well as preserve the relational structure between item embeddings, we integrate the recommendation part (i.e., $L_{rec}$) and the relation modeling part (i.e., $L_{rel}$) in an end-to-end fashion through a multi-task learning framework. The total objective function of RCF is defined as
\begin{equation}
\begin{split}
	\label{total loss}
	&\min_{\Theta}\hspace{0.3cm}L= L_{rec}+\gamma L_{rel} \hspace{0.1cm},\\
	&\hspace{0.1cm}s.t. \hspace{0.3cm} \|\mathbf{p}_u\|_2\leq1,\|\mathbf{q}_i\|_2\leq1,
	\|\mathbf{x}_t\|_2\leq1,\|\mathbf{z}_v\|_2\leq1 \\
	&\hspace{0.85cm}\forall u\in \mathcal{U},i\in \mathcal{I}, t \in \mathcal{T},v \in \mathcal{V}\\
\end{split}
\end{equation}
where $\Theta$ is the total parameter space, including all embeddings and variables of attention networks. It's obvious that both $L_{rec}$ and $L_{rel}$ can be decreased by simply scaling up the norm of corresponding embeddings. To avoid this problem during the training process, we explicitly constrain the embeddings to fall into a unit vector space. This constraint differs from traditional $L_2$ regularization which pushes parameters to the origin. It has been shown to be effective in both fields of KGE \cite{TransE,TransR} and
recommendation \cite{TransRec,mohr,tay2018latent}.  The training procedure of RCF is illustrated in Algorithm 1.

\begin{algorithm}
 \label{algorithm for RCF}
 \caption{Learning algorithm for RCF}
 	\begin{algorithmic}[1]
        \renewcommand{\algorithmicrequire}{\textbf{Input:}} 
        \renewcommand{\algorithmicensure}{\textbf{Output:}}
        \Require
        user-item interaction data $\mathcal{D}_{I}$, item relationl data $\mathcal{D}_R$, learning rate $\eta$, smoothing factor $\rho$, $\gamma$
        \Ensure
        all parameters in the learning space $\Theta$
        \State Initialize all parameters in $\Theta$
        \Repeat 
            \State Draw a mini-batch of $(u,i,k)$ from $\mathcal{D}_I$
            \State Draw a mini-batch of $(i,r,j,j^-)$ from $\mathcal{D}_R$
            \State Compute $L_{rec}$ according to Eq.(\ref{mui})-(\ref{recommendation data}) 
            \State Compute $L_{rel}$ according to Eq.(\ref{relation embedding })-(\ref{relation data})
            \State $L\leftarrow L_{rec}+\gamma L_{rel}$
            \For {each parameter $\vartheta \in \Theta$}
                \State {Compute $\partial L$/$\partial \vartheta$ on the mini-batch by back-propagation}
                \State {Update $\vartheta \leftarrow \vartheta- \eta \cdot \partial L/\partial \vartheta$}
            \EndFor
            \For {$\boldsymbol{\theta} \in \left\{\mathbf{p}_{u},\mathbf{q}_{i},\mathbf{x}_{t},\mathbf{z}_{v}  \right\}$}
                \State $\boldsymbol{\theta} \leftarrow \boldsymbol{\theta}/\max(1,\|\boldsymbol{\theta}\|_2)$
            \EndFor
        \Until converge
        \State return all parameters in $\Theta$
 	\end{algorithmic}
 \end{algorithm}
\subsection{Discussion}
Here we examine three types of related recommendation models and discuss the relationship between RCF and them.
\subsubsection{Conventional collaborative filtering}
RCF extends the item relations from the collaborative similarity to multiple and semantically meaningful relations. It can easily generalize the conventional CF methods. If we downgrade the MLP in Eq.(\ref{predicted score}) to inner product and only consider one item relation (i.e., the collaborative similarity), we can get the following predicted score:
\begin{equation}
\label{baseline}
\hat{y}_{ui}=\underbrace{\mathbf{p}_u^T\mathbf{q}_i}_{\text{MF}}+\underbrace{\mathbf{q}_i^T
\left(\sum_{j\in \mathcal{I}_u^+ \backslash \{ i \}} \beta(i,j) \cdot \mathbf{q}_{j}\right)}_{\text{NAIS}},
\end{equation}
which can be regarded as en ensemble of matrix factorization (MF) \cite{mfkoren} and the item-based NAIS model \cite{FISM}. In fact, compared with conventional CF methods, RCF captures item relations in an explicit and fine-grained level, and thus enjoys much more expressiveness to model user preference.

\subsubsection{Knowledge graph enhanced recommendation}
Recently, incorporating KG as an additional data source to enhance recommendation has become a hot research topic. These works can be categorized into embedding-based methods and path-based methods. Embedding-based methods \cite{cke,kgmemorynetwork,wang2018dkn,cao2019unifying} utilize KG to guide the representation learning. However, the central part of ICF is the item similarity and none of these methods is designed to explicitly model it. On the contrary, RCF aims at directly modeling the item similarity from both the collaborative perspective and the multiple concrete relations. Path-based methods \cite{pknn,yangjiernnpath,metapath,wang2018ripplenet,ai2018learning} first construct paths to connect users and items, then the recommendation is generated by reasoning over these paths. However, constructing paths between users and items isn't a scalable approach when the number of users and items are very large. Under such situation, sampling \cite{wang2018ripplenet,ai2018learning} and pruning \cite{pknn,yangjiernnpath} must be involved. However, RCF is free from this problem. 
Besides, the recommendation model of RCF is totally different from the path-based methods. 

\subsubsection{Relation-aware recommendation}
MCF \cite{alsoview} proposed to utilize the ``also-viewed'' relation to enhance rating prediction. However, the ``also-viewed'' relation is just a special case of the item co-interact patterns and thus still belongs to the collaborative similarity. Another work which considers heterogeneous item relations is MoHR \cite{mohr}. But it only suits for the sequential recommendation. The idea of MoHR is to predict both the next item and the next relation. The major drawback of MoHR is that it can only consider the relation between the last item of $\mathcal{I}_u^+$ and the target item. As a result, it fails to capture the long-term dependencies. On the contrary, RCF models the user preference based on all items in $\mathcal{I}_u^+$. The attention mechanism empowers RCF to be effective when capturing both long-term and short-term dependencies.
\section{Experiments}
In this section, we conduct experiments on two real-world datasets to evaluate the proposed RCF model. We aim to answer the following research questions:

\textbf{RQ1:} Compared with state-of-the-art recommendation models, how does RCF perform?
	
\textbf{RQ2:} How do the multiple item relations affect the model performance?

\textbf{RQ3:} How does RCF help to comprehend the user behaviour? Can it generate more convincing recommendation?

In the following parts, we will first present the experimental settings and then 
answer the above research questions one by one.
\subsection{Experimental Settings}
\subsubsection{Datasets} We perform experiments with two publicly accessible
datasets: MovieLens\footnote{https://grouplens.org/datasets/movielens/} and KKBox\footnote{https://www.kaggle.com/c/kkbox-music-recommendation-challenge/data}, corresponding to movie and music recommendation, respectively. Table \ref{Datasets} summarizes the statistics of the two datasets.

\textbf{1. MovieLens.} This is the stable benchmark published by GroupLens \cite{harper2016movielens}, which contains 943 users and 1,682 movies. We binarize the original user ratings to convert the dataset into implicit feedback. To introduce item relations, we combine it with the IMBD dataset\footnote{https://www.imdb.com/interfaces/}. The two datasets are linked by the titles and release dates of movies. The relation types of this data contains genres\footnote{Here, genres means that two movies share at least one same genre, as shown in Figure \ref{fig:relation-example}. Same definition also suits for the following relation types.}, directors, actors, and $t_0$, which is the relation type of the latent relation.

\textbf{2. KKBox.}  This dataset is adopted from the WSDM Cup 2018 Challenge\footnote{https://wsdm-cup-2018.kkbox.events/} and is provided by the music
streaming service KKBox. Besides the user-item interaction data, this dataset also contains description of music, which can help us to introduce the item relations. We process this dataset by removing the songs that have missing description. The final version contains 24,613 users, 61,877 items and 2,170,690 interactions. The relation types of this dataset contain genre, artist, composer, lyricist, and $t_0$.
\begin{table}
    \centering
    \setlength{\abovecaptionskip}{3pt}
    \begin{threeparttable}
    \caption{Dataset statistics.}
    \label{Datasets}
    \begin{tabular}{p{1.5cm}p{1.6cm}p{1.5cm}<{\centering}p{2.0cm}<{\centering}}
    \toprule
    &Dataset&MovieLens & KKBox\cr
    \midrule
    \multirow{3}{1.9cm}{User-Item Interactions}&\#users&943&24,613\cr
    &\#items&1,682&61,877\cr
    &\#interactions&100,000&2,170,690\cr
    \midrule
    \multirow{3}{1.9cm}{Item-Item Relations}&\#types&4&5\cr
    &\#values&5,126&42,532\cr
    &\#triplets&924,759&70,237,773\cr
    \bottomrule
  \end{tabular}
    \end{threeparttable}
    \vspace{-0.3cm}
\end{table}
\subsubsection{Evaluation protocols}
To evaluate the performance of item recommendation, we adopt the leave-one-out evaluation, which has been widely used in literature \cite{FISM,acf,he2018nais}. More precisely, for each user in MovieLens, we leave his latest two interactions for validation and test and utilize the remaining data for training. For the KKBox dataset, because of the lack of timestamps, we randomly hold out two interactions for each user as the test example and the validation example and keep the remaining for training. Because the number of items is large in this dataset, it's too time consuming to rank all items for every user. To evaluate the results more efficiently, we randomly sample 999 items which have no interaction with the target user and rank the validation and test items with respect to these 999 items. This has been widely used in many other works \cite{acf,tay2018latent,pknn,he2018nais}.

The recommendation  quality  is  measured  by  three metrics: hit ratio (HR), mean reciprocal rank (MRR) and normalized discounted cumulative gain (NDCG). HR@$k$ is a recall-based metric, measuring whether the test item is in the top-$k$ positions of the recommendation list (1 for yes and 0 otherwise).  MRR@$k$  and  NDCG@$k$  are  weighted versions which assign higher scores to the top-ranked items in the recommendation list \cite{ndcg}. 
\subsubsection{Compared methods}
We compare the performance of the proposed RCF with the following baselines:
\begin{itemize}
    \item MF \cite{bpr}: This is the standard matrix factorization which models the user preference with inner product between user and item embeddings.
	\item FISM \cite{FISM}: This is a state-of-the-art ICF model which characterizes the user with the mean aggregation of the embeddings of his interacted items.
	\item NAIS \cite{he2018nais}: This method enhances FISM through a neural attention network. It replaces the mean aggregation of FISM with an attention-based summation.
	\item FM \cite{fm}: Factorization machine is a feature-based baseline which models the user preference with feature interactions. Here we treat the auxiliary information of both datasets as additional input features. 
    \item NFM \cite{NFM}: Neural factorization machine improves FM by utilizing a MLP to model the high-order feature interactions.
    \item CKE \cite{cke}: This is an embedding-based KG-enhanced recommendation method, which integrates the item embeddings from MF and TransR \cite{TransR}.
    \item MoHR \cite{mohr}: This method is a state-of-the-art relation-aware CF method. We only report its results on the MovieLens dataset because it's designed for sequential recommendation and the KKBox dataset contains no timestamp information.
\end{itemize}
\subsubsection{Parameter settings}
To fairly compare the performance of models, we train all of them by optimizing the BPR loss (i.e.,Eq(\ref{recommendation loss})) with mini-batch Ada-grad \cite{adagrad}. The learning rate is set as 0.05 and the batch size is set as 512. The embedding size is set as 64 for all models. For all the baselines, the $L_2$ regularization coefficients are tuned between $[1e^{-5},1e^{-4},1e^{-3},0]$. For FISM, NAIS and RCF, the smoothing factor $\rho$ is set as 0.5. We pre-train NAIS with 100 iterations of FISM. For the attention-based RCF and NAIS, the attention factor is set as 32. Regarding NFM, we use FM embeddings with 100 iterations as pre-training vectors. The number of MLP layers is set as 1 with 64 neurons, which is the recommended setting of their original paper \cite{NFM}. The dropout ratio is tuned between $[0,0.1,\cdot\cdot\cdot,0.9]$. For the MLP of RCF, we adopt the same settings with NFM to guarantee a fair comparison. For MoHR, we set the multi-task learning weights as 1 and 0.1 according to their original paper \cite{mohr}. For RCF, we find that it achieves satisfactory performance when $\gamma=0.01$. We report the results under this setting if there is no special mention.
\subsection{Model Comparison (RQ1)}
\begin{table*}
    \centering
    \setlength{\abovecaptionskip}{3pt}
    \begin{threeparttable}
    \caption{Top-$k$ recommendation performance comparison of  different models ($k=5, 10, 20$). The last column RI denotes the relative improvement on average of RCF over the baseline. $*$ denotes the significance $p$-value < 0.05 compared with the best baseline on the corresponding metric (indicated by boldface).}\vspace{-5pt}
    \label{comparison between different models}
    \begin{tabular}{lccccccccc|c}
    \toprule
    \multirow{2}{*}{Models}&
    \multicolumn{9}{c}{MovieLens}&\cr
    \cmidrule(lr){2-4} \cmidrule(lr){5-7} \cmidrule{8-10} \cmidrule(lr){11-11}
    &HR@5&MRR@5&NDCG@5&HR@10&MRR@10&NDCG@10&HR@20&MRR@20&NDCG@20&RI\cr
    \midrule
    MF& 0.0774&0.0356 &0.0458 &0.1273 &0.0430 &0.0642&0.2110&0.0482&0.0833&+25.2\%\cr
    FISM&0.0795&0.0404&0.0500&0.1325 &0.0474 &0.0671 &0.2099 &0.0526&0.0865&+20.3\% \cr
    NAIS&0.0827&0.0405&0.0508&0.1367 &0.0477 &0.0683 &0.2142 &0.0528 &0.0876&+17.9\% \\ \hline
    FM&0.0827&0.0421&0.0521&0.1410& \bf 0.0496 &0.0707 &0.1994 &0.0535 &0.0852&+18.6\%\cr
    NFM & \bf 0.0880&0.0427& \bf0.0529& \bf 0.1495& 0.0495 &0.0725 &0.2153 &0.0540 & \bf 0.0889&+13.4\% \\ \hline
    CKE&0.0827&0.0414&0.0515&0.1404&0.0476&0.0688&0.2089&0.0528&0.0884&+15.2\%\cr
    MoHR&0.0832 & \bf 0.0490 &0.0499 &0.1463 &0.0485 & \bf 0.0733 & \bf 0.2249& \bf 0.0554 &0.0882&+11.2\% \\ \hline
    RCF&\bf 0.1039$^*$&\bf 0.0517$^*$&\bf 0.0646$^*$&\bf 0.1591$^*$&\bf 0.0598$^*$&\bf 0.0821$^*$ &\bf 0.2354$^*$&\bf 0.0642$^*$&\bf 0.1015$^*$\cr
    \midrule
    \multirow{2}{*}{Models}&
    \multicolumn{9}{c}{KKBox}&\cr
    \cmidrule(lr){2-4} \cmidrule(lr){5-7} \cmidrule{8-10} \cmidrule(lr){11-11}
    &HR@5&MRR@5&NDCG@5&HR@10&MRR@10&NDCG@10&HR@20&MRR@20&NDCG@20&RI\cr
    \midrule
    MF& 0.5575&0.3916 &0.4329 &0.6691 &0.4065 &0.4690&0.7686&0.4135&0.4942&+29.1\%\cr
    FISM&0.5676 &0.4084 &0.4356 &0.6866 &0.4103 &0.4844&0.7654&0.4258&0.5244&+26.2\%\cr
    NAIS&0.5862&0.4156&0.4409&0.6932&0.4153&0.4966&0.7810&0.4333&0.5315&+24.0\%\\ \hline
    FM&0.5793&0.4064&0.4495&0.6949&0.4219&0.4869& \bf 0.7941&0.4288&0.5121&+24.4\%\cr
    NFM& \bf 0.5973&0.4183& \bf 0.4630& \bf 0.7178 & \bf 0.4432& \bf 0.5088 &0.7768& \bf 0.4476&0.5244&+19.9\%\\ \hline
    CKE&0.5883& \bf 0.4191&0.4613&0.6930&0.4332&0.4952&  0.7865&0.4397& \bf 0.5389&+21.3\%\\ \hline
    RCF& \bf 0.7158$^*$ &\bf 0.5612$^*$ &\bf 0.5999$^*$ &\bf 0.7940$^*$ &\bf 0.5718$^*$ &\bf 0.6253$^*$ &\bf0.8563$^*$ &\bf 0.5762$^*$ &\bf 0.6412$^*$ \cr
    \bottomrule
    \end{tabular}
    \end{threeparttable}
\end{table*}
Table \ref{comparison between different models} demonstrates the comparison between all methods when generating top-$k$ recommendation. It's obvious that the proposed RCF achieves the best performance among all methods on both datasets regarding to all different top-$k$ values. 

Compared with the conventional item-based FISM and NAIS which only consider the collaborative similarity, our RCF is based on the multiple and concrete item relations. We argue that this is the major source of the the improvement. From this perspective, the results demonstrate the importance of multiple item relations when modeling the user preference. 

Compared with the feature-based FM and NFM, RCF still achieves significant improvement. The reason is that although FM and NFM also incorporate the auxiliary information, they fail to explicitly model the item relations based on that data. Besides, we can also see that NFM achieves better overall performance than FM because it introduces a MLP to learn high-order interaction signals. However, RCF achieves higher performance under the same MLP settings, which confirms the effectiveness of modeling item relations.

Compared with CKE, we can see that although CKE utilizes KG to guide the learning of item embeddings, it fails to directly model user preference based on multiple item relations, resulting in lower performance than RCF. Besides, we can see that although MoHR is also relation-aware, RCF still achieves better results than it. The reason is that MoHR only considers the relation between the last historical item and the target item, and thus fails to capture the long-term dependencies among the user interaction history.  
 
\subsection{Studies of Item Relations (RQ2)}
\subsubsection{Effect of the hierarchy attention}
RCF utilizes a hierarchy attention mechanism to model user preference. In this part, we conduct experiments to demonstrate the effect of the two-level attentions. Table \ref{effect of attention} shows the results of top-10 recommendation when replacing the corresponding attention with average summation. It's obvious that both the first-level and the second-level attentions are necessary to capture user preference, especially the second-level attention, which aims at calculating a specific weight for every historical item and thus largely improves the model expressiveness.
\begin{table}
    \centering
    \setlength{\abovecaptionskip}{3pt}
    \begin{threeparttable}
    \caption{Performance when replacing the attention with average summation. Avg-1 denotes the first-level attention (i.e.,$a(u,t)$) is replaced. Avg-2 means the second-level attention (i.e., $\beta_{t}(i,j,v)$) is replaced. Avg-both denotes replacing both attentions. Dec is the average decrease of performance. $*$ denotes the statistical significance for $p<0.05$.}
    \label{effect of attention}
    \begin{tabular}{lccc|c}
    \toprule
    \multirow{2}{*}{Models}&
    \multicolumn{3}{c}{MovieLens}&\cr
    \cmidrule(lr){2-4} \cmidrule(lr){5-5}
    &HR@10&MRR@10&NDCG@10&Dec\cr
    \midrule
    Avg-1 & 0.1478&0.0556 & 0.0746&-7.6\%\cr
    Avg-2 & 0.1346&0.0501 &0.0694 &-15.6\% \cr
    Avg-both&0.1294&0.0495 &0.0684 &-17.8\% \\ \hline
    RCF&\bf 0.1591$^*$&\bf 0.0598$^*$& \bf 0.0821$^*$\cr
    \midrule
    \multirow{2}{*}{Models}&
    \multicolumn{3}{c}{KKBox}&\cr
    \cmidrule(lr){2-4}\cmidrule(lr){5-5}
    &HR@10&MRR@10&NDCG@10&Dec\cr
    \midrule
    Avg-1&0.7657 &0.5484 &0.5773 &-5.0\% \cr
    Avg-2&0.6983 &0.4331 &0.5249 &-16.8\%  \cr
    Avg-both&0.6792 &0.4103 &0.4946&-20.4\% \\ \hline
    RCF& \bf0.7940$^*$& \bf 0.5718$^*$& \bf 0.6253$^*$\cr
    \bottomrule
    \end{tabular}
    \end{threeparttable}
\end{table}
\subsubsection{Ablation studies on relation modeling}
The proposed RCF defines the item relations with relation types and relation values. To demonstrate the effectiveness of these two components, we modify the proposed RCF by masking the corresponding parts. Table \ref{tab:model-modification} shows the detail of the masked models. Table \ref{results of relation ablation} reports the performance when masking different relation components. We can draw the following conclusions from this table.
\begin{enumerate}[leftmargin=*]
\item RCF-type achieves better performance than the single model, demonstrating the importance of relation types. Generally speaking, the type component describes item relations in an abstract level. It helps to model the users' preference on a class of items which share particular similarity in some macro perspectives.
\item The performance of RCF-value is also better than the single model. This finding verifies the effectiveness of relation values, which describe the relation between two specific items in a much fine-grained level. The relation value increases the model fidelity and expressiveness largely through capturing the user preference from micro perspective. 
\item RCF achieves the best performance. It demonstrates that both relation types and relation values are necessary to model the user preference. Moreover, it also confirms the effectiveness of the proposed two-level attention mechanism to tackle with the hierarchical item relations.
\end{enumerate}
\begin{table}
 \centering
 \begin{threeparttable}
 \caption{\label{tab:model-modification}Modification of RCF. Single denotes only considering one relation (i.e., collaborative similarity). RCF-type only considers relation types for the attention. RCF-value only considers relation values. $g$ denotes the attention function.}
 \setlength{\abovecaptionskip}{0.15cm}
  \begin{tabular}{p{1.5cm}|p{6.0cm}}
  \toprule
  & Modification\cr
  \midrule
  Single&Eq.(\ref{mui})$\Rightarrow \mathbf{m}_{u,i}=\mathbf{p}_u+\sum_{j \in \mathcal{I}_{u}^+}g(i,j)\cdot \mathbf{q}_j$\cr
  \specialrule{0em}{1.5pt}{1.5pt}
  RCF-type&Eq.(\ref{suit})$\Rightarrow \mathbf{s}_{u,i}^t =\sum_{j \in \mathcal{I}_{u,i}^t} g(i,j)\cdot \mathbf{q}_{j}$\cr
  \specialrule{0em}{1.5pt}{1.5pt}
  RCF-value&Eq.(\ref{mui})$\Rightarrow \mathbf{m}_{u,i}=\mathbf{p}_u+\sum_{j \in \mathcal{I}_{u}^+}g(i,j,v)\cdot \mathbf{q}_{j}$\cr
  \bottomrule
  \end{tabular}
 \end{threeparttable}
\end{table}

\begin{table}
    \centering
    \begin{threeparttable}
    \caption{Performance of different relation ablations when generating top-10 recommendation. Dec is the average decrease of performance. $*$ denotes the statistical significance for $p<0.05$.}
    \setlength{\abovecaptionskip}{0.15cm}
    \label{results of relation ablation}
    \begin{tabular}{lccc|c}
    \toprule
    \multirow{2}{*}{Ablations}&
    \multicolumn{3}{c}{MovieLens}&\cr
    \cmidrule(lr){2-4} \cmidrule(lr){5-5}
    &HR@10&MRR@10&NDCG@10&Dec\cr
    \midrule
    Single& 0.1399&0.0481 &0.0691 &-14.6\% \cr
    RCF-type &0.1484&0.0587&0.0804&-4.5\%\cr
    RCF-value &0.1548&0.0558&0.0801&-3.4\%\\ \hline
    RCF&\bf 0.1591$^*$&\bf 0.0598$^*$&\bf 0.0821$^*$\cr
    \midrule
    \multirow{2}{*}{Ablations}&
    \multicolumn{3}{c}{KKBox}&\cr
    \cmidrule(lr){2-4}\cmidrule(lr){5-5}
    &HR@10&MRR@10&NDCG@10&Dec\cr
    \midrule
    Single&0.6923&0.4666&0.5207&-15.7\% \cr
    RCF-type&0.7523&0.5431&0.5723&-6.2\%\cr
    RCF-value&0.7708&0.5579&0.5867&-3.8\% \\ \hline
    RCF&\bf 0.7940$^*$&\bf 0.5718$^*$&\bf 0.6253$^*$\cr
    \bottomrule
    \end{tabular}
    \end{threeparttable}
\end{table}
\subsubsection{Effect of multi-task learning}
RCF utilizes the item relational data in two ways: constructing the target-aware user embeddings and introducing the relational structure between item embeddings through the multi-task learning framework. In this part, we conduct experiments to show the effect of the later.

\begin{figure}
\setlength{\abovecaptionskip}{0.1cm}
\setlength{\belowcaptionskip}{-0.4cm}
    \captionsetup[subfloat]{captionskip=-5pt,nearskip=0pt,farskip=0pt}{}
    \centering
    \subfloat[MovieLens]{
    \label{gamma_ml}
    \includegraphics[width=0.25\textwidth]
    {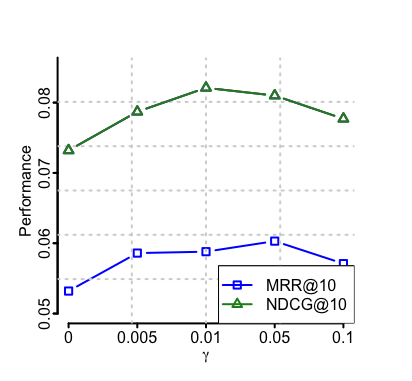}}
    \subfloat[KKBox]{%
    \label{gamma_kkbox}
    \includegraphics[width=0.25\textwidth]{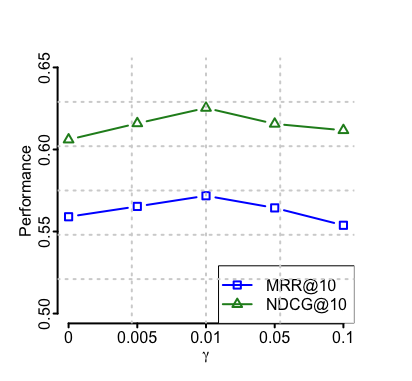}}
    \caption{Effect of $\gamma$ on the two datasets}
    \label{effect of gamma}
\end{figure}

Figure \ref{effect of gamma} reports the results of MRR@10 and NDCG@10 when changing the multi-task learning weight $\gamma$\footnote{Results of HR@10 show similar trends and are omitted due to the reason of space.}. It's obvious the performance of RCF boosts when $\gamma$ increases from 0 to positive values on both two datasets. Because $\gamma=0$ means only the recommendation task (i.e., $L_{rec}$) is considered, we can draw a conclusion that jointly training $L_{rec}$ and $L_{rel}$  can definitely improve the model performance. In fact, the function of $L_{rel}$ is to introduce a constraint that if there is relation between two items, there must be an inherent structure among their embeddings. This constraint explicitly guides the learning process of both item and relation embeddings and thus helps to improve the model performance.
We can also see that with the increase of $\gamma$, the performance improves first and then starts to decrease. Because the primary target of RCF is recommendation other than predicting item relations, we must make sure that $L_{rec}$ accounts the crucial part in the total loss. Actually, we can see from Table \ref{Datasets} that the number of item-item relational triplets is much larger than the number of user-item interactions, leading to situation that $\gamma$ is commonly set as a small value.
\subsection{Qualitative Analyses (RQ3)}
In this part, we conduct qualitative analyses to show how RCF helps us to comprehend user behaviors and generate more convincing recommendation.
\subsubsection{Users as a whole}
\begin{figure}
\setlength{\abovecaptionskip}{0.1cm}
\setlength{\belowcaptionskip}{0cm}
    \centering
    \captionsetup[subfloat]{captionskip=0pt,nearskip=0pt,farskip=0pt}
    \centering
    \subfloat[MovieLens]{
    \includegraphics[width=0.25\textwidth]{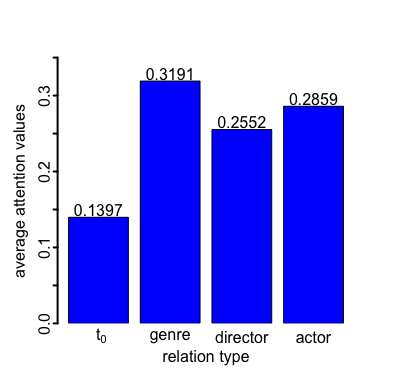}}
    \subfloat[KKBox]{%
    \includegraphics[width=0.25\textwidth]{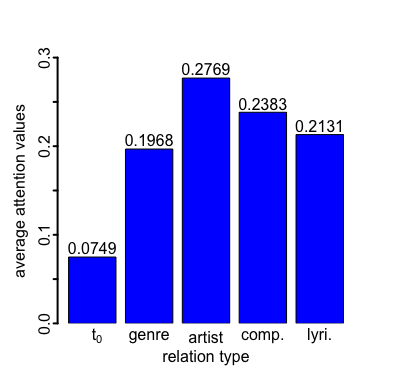}}
    \caption{Average $a(u,t)$ on two datasets. $a(u,t)$ denotes the user $u$'s attention on the relation type $t$.}
    \label{user's attention on types}
\end{figure}

Figure \ref{user's attention on types} illustrates the average $a(u,t)$ for all $u \in \mathcal{U}$ on the two datasets. We can see that on the MovieLens dataset, the largest $a(u,t)$ falls into genre, which means that users tend to watch movies that share same genres. The second position is actor. This finding is in concert with the common sense that genres and actors are the most two important elements that affect the users' choices on movies. Director is in the third position. Moreover, we can see that all these three relation types are more important than $t_0$, which denotes the collaborative similarity. It further confirms that only considering collaborative similarity is not enough to model user preference. Multiple and fine-grained item relations should be involved to generate better recommendation.

For the music domain, we can see that the most important relation type falls into artist. Following that are comp. (short for composer) and lyri. (short for lyricist). They are the most three important factors that affect users when listening to music. Besides, compared with the movie domain, the attention $a(u,t_0)$ in the music domain is much smaller. It indicates that user behaviour patterns when listening to music are more explicit than the ones when watching movies. As a result, our proposed RCF achieves bigger improvement on the KKBox dataset, as shown in Table \ref{comparison between different models}. 

\subsubsection{Invididual case studies}
\begin{figure}
\setlength{\abovecaptionskip}{0.1cm}
\setlength{\belowcaptionskip}{0cm}
	\centering
    \includegraphics[width=0.45\textwidth]{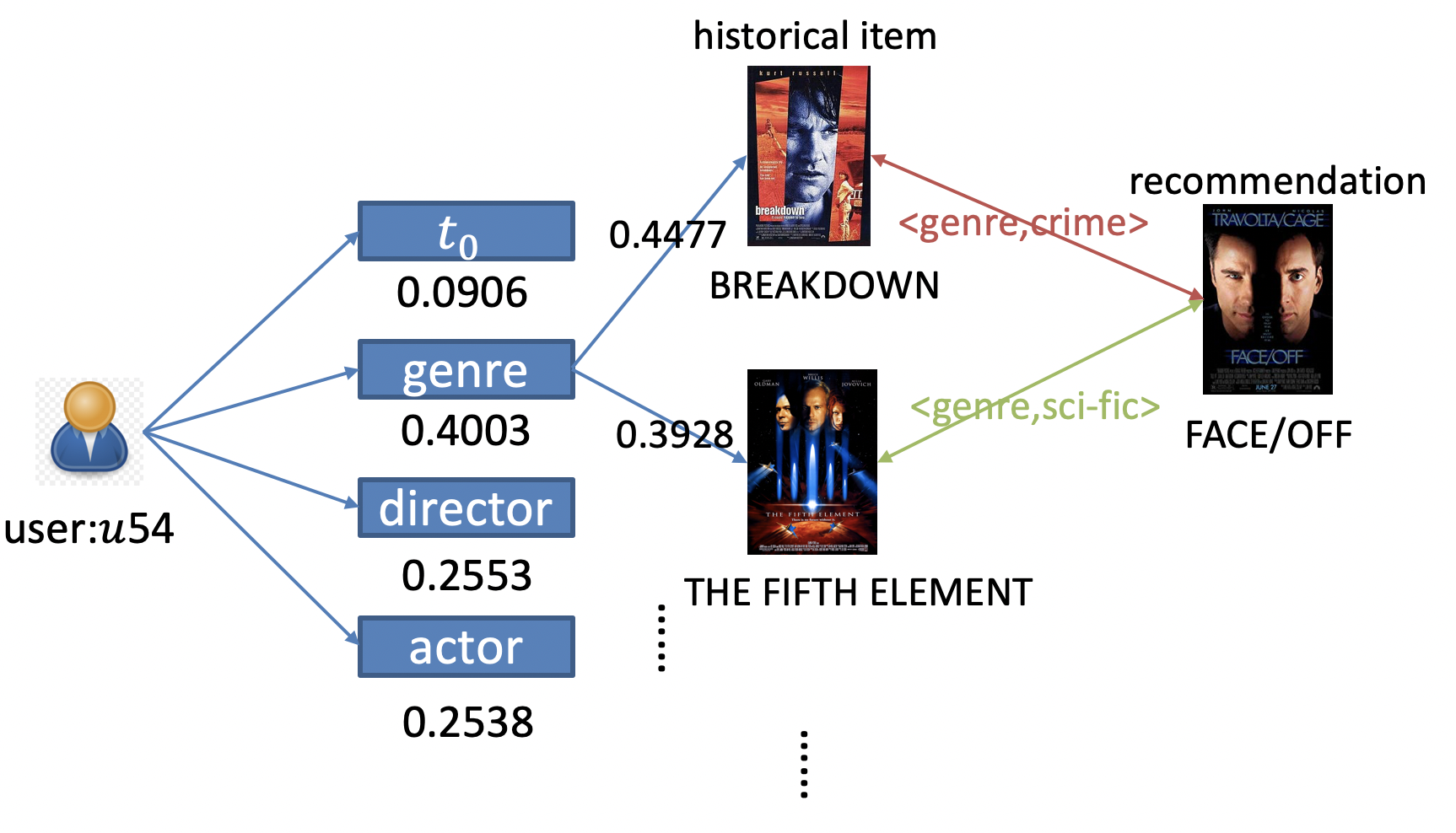}
    \caption{Attention visualization of user $u54$ in MovieLens. } 
    \label{explaination}
\end{figure}

We randomly select a user $u54$ in the MovieLens dataset to see how RCF helps us to comprehend the individual user behavior. Figure \ref{explaination} shows the attention visualization of this user. We can see that this user pays the most attention (0.4003) on the relation type ``shared genres'' when watching movies. Among the second-level relation values, he is most interested in ``crime'' (0.4477) and ``sci-fic'' (0.3928). Based on his historical interacted movies ``Breakdown'' and ``The Fifth Element'', the recommended movie is ``Face/Off''. From this perspective, we can also generate the explanation as \emph{``Face/Off'' is recommended to you because it is a crime movie like ``Breakdown'' you have watched before}. It's obvious that a side benefit of RCF is that it can generate reasonable explanations for recommendation results. 

\section{Related Work}
\subsection{Item-based Collaborative Filtering}
The idea of ICF is that the user preference on a target item $i$ can be inferred from the similarity of $i$ to all items the user has interacted in the past \cite{itemCF,linden2003amazonitemCF,he2018nais,FISM}. Under this case, the relation between items is referred as the collaborative similarity, which measures the co-occurrence in the user interaction history.
A popular approach of ICF is FISM \cite{FISM}, which characterizes the user representation as the mean aggregation of item embeddings which occur in his interaction history. Plenty of work has been done following this research line, such as incorporating user information \cite{elbadrawy2015userspecific,xin2016fhsm}, neural network-enhanced approaches \cite{wu2016collaborativeautoencoder,he2018nais,he2018apr} and involving local latent space \cite{christakopoulou2018localsvd,lee2013local}.

Although these methods has improved the performance of ICF, all of them are based solely on the collaborative similarity between items. This item relation is coarse-grained and lacks of semantic meaning, introducing the bottleneck of the model and the difficulty of generating convincing results.

\subsection{Attention Mechanism}
The attention mechanism has become very popular in fields of computer vision \cite{attentioncv1,attentioncv2} and natural language processing \cite{attnention,selfattention} because of its good performance and interpretability for deep learning models. The key insight of attention is that human tends to pay different weights to different parts of the whole perception space.
Based on this motivation, \cite{he2018nais} improved FISM by replacing the mean aggregation with attention-based summation. \cite{acf} proposed to utilize the attention mechanism to generate multimedia recommendation. \cite{kang2018selfattentive} exploited self-attention for sequential recommendation. 
There are many other works focusing on involving attention mechanism for better recommendation \cite{afm,tay2018latent}. However, all of them fail to model the multiple item relations. In fact, users tend to pay different weights on different item relations and it's a promising direction to utilize attention mechanism under such circumstance. 

\section{Conclusion}
In this work, we proposed a novel ICF framework namely RCF to model the multiple item relations for better recommendation. RCF extends the item relations of ICF from collaborative similarity to fine-grained and concrete relations. We found that both the relation type and the relation value are crucial for capturing user preference.  Based on that, we proposed to utilize a hierarchy attention mechanism to construct user representations. Besides, to maximize the usage of relational data, we further defined another task which aims to preserve the relational structure between item embeddings. We jointly optimize it with the recommendation task in an end-to-end fashion through a multi-task learning framework. Extensive experiments on two real-world datasets show that RCF achieves significantly improvement over state-of-the-art baselines. Moreover, RCF also provides us an approach to better comprehend user behaviors and generate more convincing recommendation. 

Future work includes deploying RCF on datasts with more complex item relations. Besides, we also want to extend RCF to empower it with the ability to tackle with not only the static item relations but also the dynamic user relations. Another promising direction is how to utilize item relations to develop adaptive samplers for the pair-wise ranking framework.

\vspace{0.3cm}
\noindent\textbf{Acknowledgements}. The GPUs used in this research are provided by NVIDIA Corporation with 1080Ti series. This research is supported by the Thousand Youth Talents Program 2018. Joemon Jose and Xiangnan He are corresponding authors.

\bibliographystyle{ACM-Reference-Format}
\bibliography{sample-bibliography}

\end{document}